\documentstyle[12pt]{article}
\newenvironment{namelist}[1]{%
\begin{list}{}
{

\settowidth{\labelwidth}{#1}
\setlength{\leftmargin}{1.05\labelwidth}
}
}{%
\end{list}}
\newcommand{\dis}{\displaystyle}
\begin{document}
\def\qed{~\vrule height6pt width4pt depth0pt\medskip}
\setlength{\baselineskip}{0.35in}
\setlength{\jot}{0.2in}

\begin{center}
{\large \bf
The velocity and angular momentum of a free Dirac electron
}\\

\vspace*{0.5cm}
Lu Lin
\\
Department of Electrophysics\\
National Chiao Tung University\\
Hsinchu, Taiwan \\
Republic of China\\
\end{center}

\vspace*{2cm}

\noindent
{\bf ABSTRACT}~~\\

It is shown that, in Dirac theory, there is a spatial velocity of a free electron
which commutes with the Hamiltonian, so it is a conserved quantity
of the motion.  Furthermore, there is a spatial orbital angular momentum
which also commutes with the Hamiltonian and is a constant of the motion.\\
\newpage

\noindent
1. The spatial velocity for a of Dirac electron:\\

To descrive the motion of a free electron, besides the space-time
coordinates $(\mbox{\boldmath $X$}, ct)=(x_{1}x_{2}x_{3}x_{4})$, Dirac
introduced also $(\alpha _{1}\alpha_{2}\alpha_{3}\beta )
=(\mbox{\boldmath $\alpha$}
\beta )$ for the internal motion in the electron which operate
on a complex spinor space.  Since the spatial space and the
spinor space are basically independent, so the $\alpha 's$ and $\beta $
are independent of and commute with the $x's$ and the $p's$, where
$(p_{1}p_{2}p_{3}p_{4})$ are the momentum and energy of the electron.
The Hamiltonian of the free electron is gien as
\begin{eqnarray}
H=c\mbox{\boldmath $\alpha$}\cdot \mbox{\boldmath $P$}
+mc^{2}\beta .
\end{eqnarray}

\noindent
Dirac showed that
\begin{eqnarray}
& & \frac{d}{dt}\mbox{\boldmath $X$} = c\mbox{\boldmath $\alpha$},\\
& & \mbox{\boldmath $X=\xi$}+\mbox{\boldmath $\eta$},\\
& & \mbox{\boldmath $\xi =\xi$}_{0} +c^{2}\mbox{\boldmath $P$}
H^{-1}t,\\
& & \mbox{\boldmath $\eta$}=\frac{i}{2}c\hbar (\mbox{\boldmath $\alpha $}
H^{-1}-c\mbox{\boldmath $P$}H^{-2})
\end{eqnarray}

\noindent
where $\mbox{\boldmath $\xi$}_{0}$
is the $\mbox{\boldmath $\xi$}(t)$
when $t=0$, and the $\sigma 's$ the Pauli matrices.
Next, we calculate the time
derivative of $\mbox{\boldmath $\eta $}$ by taking the
commutator of $\mbox{\boldmath $\eta $}$ with $H$
and obtain
\begin{eqnarray}
\frac{d}{dt}\mbox{\boldmath $\eta$}=c\mbox{\boldmath $\alpha$}-c^{2}
\mbox{\boldmath $P$}H^{-1}.
\end{eqnarray}

\noindent
From equation (2) and equation (6), we get
\begin{eqnarray}
\frac{d}{dt}\mbox{\boldmath $\xi$}=\frac{d}{dt}
(\mbox{\boldmath $X-\eta$})=c^{2}\mbox{\boldmath $P$}H^{-1}.
\end{eqnarray}

\noindent
The operator
$c^{2}\mbox{\boldmath $P$}H^{-1}$
commutes with the Hamiltonian $H$, and has eigen value
$\mbox{\boldmath $v$}=\mbox{\boldmath $P$}/M $ ($M$ is the total mass),
which is a velocity vector of a free particle.
So it can be
simultaneously measured with $H$ and is a conserved quantity during the
course motion.  Therefore we conclude that $c^{2}P_{i}H^{-1}=v_{i}$ is
the eigen value obtained by a measurement of the i-component of the
velocity of a free electron.\\

It may be worthwhile to mention that, since for the moment we
do not know exactly the operation rules in the theory for the vector
$\mbox{\boldmath $\xi$}$.  So when we have to deal
with this vector, we should express it as
$\xi_{i}=x_{i}-\eta_{i}$, and then deal with $x_{i}$ and
$\eta_{i}$.  Otherwise we may have trouble.  For instance, the
commutator for $p$ and $q$ is $[q,p]=i\hbar$, so
$[x_{1},p_{1}]=i\hbar=[\xi_{1}+\eta_{1},p_{1}]=[\xi_{1},p_{1}]
+[\eta_{1},p_{1}]$.  From equation (5) we see that $\eta_{1}$
commutes with $p_{1}$ so $[\eta_{1},p_{1}]=0$, and
$[\xi , p_{1}]=i\hbar$.  Then if we take $[x_{1},H]
=[\xi_{1},H]+[\eta_{1},H]=i\hbar \dot{x}_{1}$, we will get
$\dot{x}_{1}=c\alpha_{1}+c\alpha_{1}-c^{2}p_{1}H^{-1}
=2c\alpha_{1}-c^{2}p_{1}H^{-1}$, which is wild when we compare it with
equation (2).  This example says that we can not operate
directly with the $\xi 's$.  We have to operate with the
$(x's-\eta 's)$.\\

Consider two classical harmonic oscillators with coordinates $z_{1}$
and $z_{2}$ and their Hamiltonions $H_{1}$ and $H_{2}$.  When
there is no coupling between them, $z_{1}$ and $z_{2}$
will oscillate independently and seperately.  However, if there  is
a coupling between them, the total Hamiltonian will become
$H_{1}+H_{2}+H_{12}$.  When we diagonalize the system, the
coordinates $z_{1}$ and $z_{2}$ will get mixed, and the normal
coordinates will be a mixture of $z_{1}$ and $z_{2}$.\\

If $\Psi$ is an eigenfunction of the Hamiltonian $H$ in equation
(1), then it is an eigenfunction of the Hamiltonion
$H'=H^{2}$.  The Hamiltonian $H'$ has only spatial
quantities (without $\mbox{\boldmath $\alpha $}$ and $\beta $).
Thus $\Psi$ must be able to describe a system which has only spatial
variables. On the other hand, $H$ is a Hamiltonian which contains
interactions between momentum and spin.  So the generalized coordinates
of the system (the $x_{i}'s$), which are the dynamical conjugate
variables of momentum, must be a mixture of both spatial and spin
variables as we can see from equations (3,4,5).  The time derivatives of
the $x_{i}'s$ are then mixtures of the time derivatives of both kinds
of variables.  However, only the component which is pure spatial is
experimentally measurable.\\

\noindent
2. The spatial orbital angular momentum of a free Dirac electron:\\

Consider an angular momentum $\mbox{\boldmath $\ell$}$ defined as
\begin{eqnarray}
\mbox{\boldmath $\ell$}=(\mbox{\boldmath $X$}-\mbox{\boldmath $\eta$}
)\times \mbox{\boldmath $P$}=\mbox{\boldmath $\xi$}\times
\mbox{\boldmath $P$}.
\end{eqnarray}

\noindent
Since $\mbox{\boldmath $P$}$ is a constant vector of the motion,
from equation (7) we see that the time derivative of $\mbox{\boldmath $\ell$}$
vanishes.  So $\mbox{\boldmath $\ell$}$ commutes with $H$ and
is a constant of the
motion and can be simultaneously measured with the Hamiltonian.
Also, $v_{i}$ is a spatial velocity and is a constant of motion.
$\xi_{i}$ is the time integration of $v_{i}$, so it is a spatial
variable.  Therefore, $\mbox{\boldmath $\ell$}$ is the spatial angular
momentum of the free electron.\\

It is well known that the electron has an intrinsic angular momentum of
$\dis\frac{\hbar}{2}\mbox{\boldmath $\sigma$}$ due to the $SU(2)$ symmetry.  In reference
(Dirac 1958),
Dirac showed that the $\sigma 's$ do not commute with the Hamiltonian (1),
so the spin angular momentum is not a constant of the motion.
But the Hamiltonian (1) is rotational invariant, the total angular momentum
must be conserved.  Through the interaction between the spin and the spatial
variables, the rotational symmetry will induce an angular momentum term,
which arises from coupling the two parts of variables in order to
conserve the total angular momentum.  The simplest form of this term which
meets those requirements is $\mbox{\boldmath $\eta$}\times \mbox{\boldmath $P$}$.
Finally we have for the total angular momentum as
\begin{eqnarray}
\mbox{\boldmath $j$}=\mbox{\boldmath $\xi$}\times \mbox{\boldmath $P$}
+\mbox{\boldmath $\eta$}\times \mbox{\boldmath $P$}
+\frac{\hbar}{2}\mbox{\boldmath $\sigma$}.
\end{eqnarray}

\noindent
If we combine the first two terms together, we will just have
$\mbox{\boldmath $j=X$}\times \mbox{\boldmath $P$}+\dis\frac{\hbar}{2}
\mbox{\boldmath $\sigma$}$.  However, equation (9) states that the total
angular momentum is a sum of a spin, a spatial, and a spin-spatial
coupling term.\\

The author wishes to thank Prof. Y.S. Wu for valuable discussions.\\

\noindent
Reference
\begin{namelist}{Dirac PAM 1958xx}
\item [{Dirac PAM 1958}]
Dirac PAM 1958 The principles of quantum mechanics, fourth ed. Oxford
Univ. Press, London.
\end{namelist}
\end{document}